\documentclass[preprint,amsmath,amssymb,dvips,pra,showpacs]{revtex4}
\usepackage{graphicx}
\usepackage{bm}
\begin{document}
\title{Density of defects and the scaling law of the entanglement entropy in quantum phase transition of one dimensional spin systems induced by a quench}
\author{Banasri Basu}
\email{banasri@isical.ac.in}
\affiliation{Physics and Applied Mathematics Unit, Indian Statistical Institute, Kolkata 700 108, India}
\author{Pratul Bandyopadhyay} 
\email{b_pratul@yahoo.co.in}
\affiliation{Physics and Applied Mathematics Unit, Indian Statistical Institute, Kolkata 700 108, India}

\author{Priyadarshi Majumdar}
\email{majumdar_priyadarshi@yahoo.com}
\affiliation{Jyotinagar Bidyasree Niketan H.S. School, 41 Jyotinagar, Kolkata 700 108, India}

\vskip 2 cm

\begin{abstract}

\noindent We have studied quantum phase transition induced by a
quench in different one dimensional spin systems. Our analysis is
based on the dynamical mechanism which envisages nonadiabaticity in
the vicinity of the critical point. This causes spin fluctuation
which leads to the random fluctuation of the Berry phase factor
acquired by a spin state when the ground state of the system evolves
in a closed path. The two-point correlation of this phase factor is
associated with the probability of the formation of defects. In this
framework, we have estimated the density of defects produced in several
one dimensional spin chains. At the critical region, the entanglement entropy of a block of $L$ spins with the rest of the system is also estimated which is found to increase
logarithmically with $L$. The dependence on the quench time puts a constraint on the block size $L$. It is also pointed out that the Lipkin-Meshkov-Glick model in point-splitting regularized form appears as a combination of the XXX model and Ising model with magnetic field in the negative z-axis.
This unveils the underlying conformal symmetry at criticality which
is lost in the sharp point limit. Our analysis shows that the
density of defects as well as the scaling behavior of the
entanglement entropy follows a universal behavior in all these
systems.


\end{abstract}


\maketitle


\section{Introduction} \label{sec1}

\noindent Kibble~\cite{1} has pointed out that in second order phase
transitions if the system is driven at a fixed rate characterized by a quench time its
evolution cannot be adiabatic close to the critical point. The
nonadiabatic evolution in the critical region produces defects such
that the system becomes a mosaic of ordered domains whose finite
size depends on the transition rate. In the cosmological scenario
Kibble considered relativistic causality to set the size of the
domains. Later, Zurek~\cite{2} suggested the dynamical mechanism
based on the universality of critical slowing down and predicted
that the size of the ordered domains scales with the transition time
$\tau_q$ as $\tau_q^{\chi}$ where $\chi$ is a critical exponent. The
Kibble-Zurek mechanism (KZM) in second order phase transition
essentially involves thermal fluctuation which initiates symmetry
breaking. In recent times, the zero temperature quantum phase
transition (QPT) has been studied in the light of KZM and the
formation of defects in one dimensional transverse Ising model has
been investigated in details by several authors~\cite{3,4,5}. In QPT
of the spin chains, apart from the study of the formation of defects
it is also important to examine  how entangled various parts of the
system are with each other. Recently, it has been shown that for
a quantum spin system in a mixed state the entanglement of formation
of a pair of nearest neighbor spins $i.e.$ concurrence is related to
the Berry phase factor acquired by a spin state when the ground
state of the system evolves in a closed path~\cite{5a,6,7}. When a
bipartite quantum system is in a pure state the measure of
entanglement between two subsystems is given by Von-Neumann entropy.
When the bipartite system is in a mixed state the entanglement of
formation given by concurrence has the property that it reduces to
the Von-Neumann entropy in a pure state~\cite{8}. It has been
observed that the entanglement entropy of a block of $L$ spins with
the rest of the system at criticality follows a logarithmic scaling
law in one-dimensional spin systems. The basic ingredient behind
this result is that at criticality the correlation length 
diverges and the system becomes conformal invariant. This conformal
symmetry leads to the logarithmic scaling law of the entanglement
entropy at criticality~\cite{9}. When the criticality is induced by
a quench the scaling behavior changes. The formation of defects as
well as the entanglement entropy of a block of $L$ spins with the
rest of the system in transverse Ising model, when QPT is induced by
a quench, has been studied in details by Cincio \textit{et
al}~\cite{5}.
\\
\noindent The behavior of the geometric phase in XY spin chain
during a linear quench caused by a gradual turning off of the
magnetic field has been studied earlier~\cite{9a}. In a recent
note~\cite{10} we have formulated a dynamical mechanism for QPT
induced by a quench based on the fact that spin fluctuation in the
vicinity of the critical point leads to the fluctuation of the Berry
phase acquired by the ground state of the system when it evolves in
a closed path. The probability of the formation of defects is
determined by the two-point correlation of this geometric phase
factor. We have studied the formation of defects in transverse Ising
model during critical slowing down based on this formalism and the
results are found to be identical with those derived by other
authors~\cite{3,4,5}. We have also studied the scaling law of the
entanglement entropy in the transverse Ising model at criticality
induced by a quench and it is shown that the prefactor now depends
on the quench time and there is a restriction on the block size
$L$~\cite{11}. In this paper,  we study the formation of defects as
well as the scaling law of the entanglement entropy at criticality
induced by a quench in some other one dimensional spin
systems on the basis of this dynamical mechanism of QPT induced by a
quench.
\\
\noindent The organization of the paper is as follows. In
sec.\ref{sec2} we briefly  review of our work~\cite{10} related to
the isotropic XY model (transverse Ising model) for comprehension of
the formulation we have adopted. In sec.\ref{sec3} and
sec.\ref{sec4} we study the XX and XXX model respectively. In
sec.\ref{sec5} we deal with the Lipkin-Meshkov-Glick (LMG) model.
 We make a comparative analysis of the results
obtained for different spin systems in sec.\ref{sec6}. Finally, we
discuss results in sec.\ref{sec7}.

\vskip .5 cm

\section{The XY model} \label{sec2}

\noindent The Hamiltonian for the XY model is given by
\begin{equation}
H_{XY}=-\sum_{i}\left(\frac{1+\gamma}{2}\sigma_i^x\sigma_{i+1}^x+\frac{1-\gamma}{2}\sigma_i^y\sigma_{i+1}^y+\lambda \sigma_i^z\right). \label{eq1}
\end{equation}
Here $\lambda$ is the external magnetic field, $\gamma$ is the
anisotropy parameter and $\sigma_i^{a}(a=x,y,z)$ are the Pauli
matrices. The XY model with ($\gamma \neq 0$) falls into the free
fermion universality class and is critical for $\lambda=1$. The
isotropic $XY$ model ($\gamma=1$) corresponds to the transverse
Ising model. We consider the Hamiltonian for the transverse Ising
model
\begin{equation}
H=-\sum_{i}\left(\sigma_i^z\sigma_{i+1}^z+\lambda \sigma_i^x\right). \label{eq2}
\end{equation}
For $\lambda >>1$ all the spins are aligned along the x-axis and the
system is in the paramagnetic state. In the region $\lambda<<1$ the
system is in the ferromagnetic state with all spins either up or
down. During critical slowing down we introduce a linear quench
\begin{equation}
\lambda(t<0)=-\frac{t}{\tau_q}, \label{eq3}
\end{equation}
where $\tau_q$ is the quench time. In the vicinity of the critical
region, the nonadiabaticity induces spin fluctuation. This
eventually causes fluctuation of the Berry phase acquired by a spin
state when the ground state evolves in a closed path as the Berry
phase acquired by a spin state is given by $\Gamma=\pi(1-cos\theta)$
where $\theta$ is the angle of deviation of the spin axis from the
quantization axis~\cite{12,13,14,15}. In the critical region, due to
nonadiabaticity some spins get excited and defects (kinks) are
formed. The Berry phase factor associated with such a spin is given
by
\begin{equation}
\phi_{k_0}=\frac{\Gamma_{k_0}}{2\pi}=\frac{1}{2}\left(1-cos\theta_{k_0}\right), \label{eq4}
\end{equation}
where $k_0$ is the momentum mode of the quasiparticle corresponding
to the spin undergoing excitation and $\theta_{k_0}$ is the
corresponding angle of deviation of the spin axis from the $z$-axis.
In our formulation it is assumed that during criticality
$\phi_{k_0}$ undergoes stochastic fluctuation and it follows the
simplest stochastic differential equation
\begin{equation}
d\phi_{k_0}(t)=-\omega_{k_0}\phi_{k_0}(t)dt+d\eta (t), \label{eq5}
\end{equation}
where $\omega_{k_0}$ is the frequency related to the energy $\epsilon_{k_0}$ near criticality and $\eta(t)$ is a Gaussian white noise satisfying the moments
\begin{eqnarray}
\langle d\eta(t) \rangle=0, \nonumber \\
\langle d\eta(t) d\eta(t')\rangle=\delta(t-t')dt'. \label{eq6}
\end{eqnarray}
This gives rise to the correlation
\begin{eqnarray}
\langle \phi_{k_0}(t) \rangle=0, \nonumber \\
\langle \phi_{k_0}(t) \phi_{k_0}(t')\rangle=\frac{1}{2}e^{-\omega_{k_0}(t-t')}. \label{eq7}
\end{eqnarray}
At criticality during the quench time $\tau_q$, the spin states
transit from the paramagnetic state to the ferromagnetic state so
that the angle of deviation ($\theta_{k_0}(\tau_q)$) of a spin axis
is $\pi/2$. The spin state undergoing excitation settles down at
$t=0$ with the spin axis reversed so that $\theta_{k_0}(0)$ takes
the value $\pi$ with certain probability $p_{k_0}$. This implies
that the random variable $\phi_{k_0}(t)$  acquires the value $1/2$
and $1$  at $t=\tau_q$ and $t=0$ respectively $i.e$, we have
$\phi_{k_0}(\tau_q)=\frac{1}{2}$, $\phi_{k_0}(0)=1$. This suggests
that the random variable $2\phi_{k_0}(\tau_q)\phi_{k_0}(0)$ attains
the value 1 with probability $p_{k_0}$. Thus using eqn.\eqref{eq7}
the excitation probability $p_{k_0}$ can be written as
\begin{equation}
p_{k_0}=2\langle \phi_{k_0}(\tau_q)\phi_{k_0}(0)\rangle=e^{-\omega_{k_0}\tau_q}. \label{eq8}
\end{equation}
The energy $\epsilon_{k_0}$ near the critical point is given by~\cite{10}
\begin{equation}
\epsilon_{k_0}=2(1-cos~k_0)=4sin^2\frac{k_0}{2} \sim k_0^2, \label{eq9}
\end{equation}
for small $k_0$. This gives
\begin{equation}
p_{k_0}=e^{-2\pi k_0^2\tau_q}, \label{eq10}
\end{equation}
The number density of defects is given by
\begin{equation}
n_1=\frac{1}{2\pi}\int_{-\pi}^{\pi}p_{k_0}dk_0=\frac{1}{2\pi}\int_{-\pi}^{\pi}e^{-2\pi\tau_q k_0^2} dk_0=\frac{1}{2\pi}\frac{1}{\sqrt{2\tau_q}}. \label{eq11}
\end{equation}
This gives rise to the domain size corresponding to the Kibble-Zurek
(KZ) correlation length $\widehat{\xi}\sim \sqrt{\tau_q}$. This
result is identical with that obtained by other
authors~\cite{3,4,5}.
\\
\noindent At the critical point the ground states are characterized
by the fact that the entanglement entropy $S_L$ of a block of $L$
spins with the rest of the system diverges like $log~L$ with a
prefactor determined by the central charge $c$ of the relevant
conformal field theory ~\cite{9}. In fact,
\begin{equation}
S_L \approx \frac{c}{3}~log_2L. \label{eq12}
\end{equation}
For the critical Ising model $c=1/2$,
\begin{equation}
S_L \approx \frac{1}{6}~log_2L,  \label{eq13}
\end{equation}
and the system falls into the free fermion universality class. It
has been shown that the central charge $c$ is related to the Berry
phase factor $\phi$, the phase being $e^{i2\pi \phi}$ and just like
the central charge $c$, the Berry phase factor $\phi$ also undergoes
a renormalization group (RG) flow~\cite{16} such that 
$L\frac{\partial \phi}{\partial L} \leq 0$ where
$L$ is a length scale. From this we have
\begin{equation}
|\phi|_L=a~ln~L=\bar{a}~log_2L.  \label{eq14}
\end{equation}
where $a (\leq 0)$ is a parameter. 
\\
\noindent For a pair of nearest neighbor spins ($L=2$), $\bar{a}=|\phi|$ which
corresponds to the entanglement of formation given by concurrence in
a mixed state~\cite{6,7}. So for the entropy of a block of $L$ spins
due to entanglement with the rest of the system in the pure state we
can write
\begin{equation}
S_L \approx |\phi|~log_2L.  \label{eq15}
\end{equation}
At criticality the concurrence for the entanglement of a pair of
nearest neighbor spins in transverse Ising model is given by
$|\phi|=0.18$~\cite{6} and so we find
\begin{equation}
S_L \approx 0.18~log_2L. \label{eq16}
\end{equation}
It is observed that the prefactor in eqn.\eqref{eq13} is very close
to the value 0.18. We may mention here that the entanglement of a
block of $L$ spins with the rest of the system can be considered to
be equivalent to the entanglement between a single spin representing
the block spin with another spin represented by the rest of the
system in block variable RG scheme. In view of this, $S_L$ in
eqn.\eqref{eq13} can be considered as the concurrence $C$ for the entanglement between the
pair of this two block variable renormalized spins in a mixed state.
The slight departure of the prefactor $\frac{1}{6}$ in
eqn.\eqref{eq13} from the value 0.18 in eqn.\eqref{eq16} may be
associated with the block variable renormalization of the spin
system which induces a change in the coupling constant. The factor
$\frac{1/6}{0.18}=0.926$ is considered to be the correction factor
associated with the block spin variable.

\noindent The introduction of a quench incorporates a new length
scale given by the KZ correlation length $\widehat{\xi}\sim
\sqrt{\tau_q}$. So to evaluate the entanglement entropy immediately
after the dynamical phase transition we have to rescale the Berry
phase factor $|\phi|$ in expression \eqref{eq15} corresponding to
the entanglement of a pair of nearest neighbor spins in terms of
that of the block spins in the domain $\widehat{\xi}$ with the rest
of the system. This follows from the fact that after the formation
of defects the system in the final state represents a kink-antikink
chain with lattice space approximately given by $\widehat{\xi}$. So
we write~\cite{11}
\begin{equation}
S(L,\tau_q)=2\frac{|\phi|~log_2L}{|\phi|~log_2\widehat{\xi}}\times 0.926 \approx 3.7 \frac{ln~L}{ln~\tau_q}. \label{eq17}
\end{equation}
The factor 0.926 has been introduced as the correction factor of the
block variable renormalization as discussed above. Eqn.(17) shows
that in the scaling law of the entanglement entropy there is a
prefactor depending on $\tau_q$
This imposes a restriction on the maximum value of $L,~L_{max}$
which is allowed in the system. As in the final state the system
represents a kink-antikink chain with lattice constant
$\widehat{\xi}$, the maximum value of the entanglement entropy
attained by the system corresponds to the entanglement of a block
$\widehat{\xi}$ spins with the rest of the system along with the
entropy associated with the entanglement of kink-antikink pair. It
As we know for a bipartite spin $1/2$ system the maximum value of
entanglement entropy is 1, which at criticality is distributed over
the whole chain, we can write for the entire chain~\cite{11}
\begin{equation}
S_{max}=2(0.18~log_2\widehat{\xi}+1)\times 0.926=0.12~ln~\tau_q+1.85.  \label{eq18}
\end{equation}
This suggests the relation
$\frac{S(L,\tau_q)}{S_{max}}\leq 1$ which implies~\cite{17}
\begin{equation}
ln~L \leq 0.03~(ln \tau_q)^2+0.5~ln~\tau_q.  \label{eq19}
\end{equation}
This restricts the block size $L$ when QPT is induced by a
quench and suggests that for sufficiently small $\tau_q$ the block
size $L \leq \widehat{\xi}$. Thus we find that during critical
slowing down due to the formation of defects, the scaling law for
the entanglement entropy is valid only in a restrictive sense.
\\
We have compared our result with that obtained by Cincio \textit{et. al.}~\cite{5} using standard technique of density matrix analysis in \cite{11}. Within the limited range of the block size $L$ determined by eqn.\eqref{eq19} the result was found to be in good agreement with that of Cincio \textit{et. al.}.

\vskip .5cm

\section{The XX model} \label{sec3}

\noindent The XX model given by the Hamiltonian
\begin{equation}
H=-\sum_i\left(\sigma_i^x\sigma_{i+1}^x+\sigma_i^y\sigma_{i+1}^y\right)+\lambda \sum_i \sigma_i^z \label{eq20}
\end{equation}
has two limit behavior. At $\lambda=2$ the system corresponds to a ferromagnetic state while at $\lambda=0$ the system falls into the free boson universality class. The interval between these two points corresponds to the critical region. As the system transits through the point $\lambda=2$ in the vicinity of the critical point we introduce a quench so that the time dependent magnetic field is given by
\begin{equation}
\lambda(t<0)=-2t/\tau_q, \label{eq21}
\end{equation}
$\tau_q$ being the quench time. At $t=0~~(\lambda=0)$ the system
settles down in a free bosonic state when the ground state is
ordered in the xy plane and we have pairs of spins with opposite
orientations. However, due to nonadiabaticity in the transition,
this simple bosonic picture is disturbed and the system is
characterized by domains having spins with same orientations i.e.
defects (kinks) are  formed. Following the analysis in the previous
section, in the critical region we consider the fluctuation of the
Berry phase acquired by a spin state when the ground state evolves
in a closed path. Let the system initially be in the the region
$\lambda>2$. As the system transits through the point $\lambda=2$,
the ferromagnetic order is  destroyed when certain spins are
flipped. From eqn.\eqref{eq4} we can easily show that the Berry
phase factor $\phi_{k_0}(\tau_q)$ takes the value 1 corresponding to
the value $\theta_{k_0}(\tau_q)=\pi$ where $k_0$ is the momentum
mode of the quasiparticle associated with the flipped spin.
\\
\noindent As time evolves, in the vicinity of $\lambda=0$ due to
nonadiabaticity certain spins get excited with probability $p_{k_0}$
and their orientations are reversed. Finally at $\lambda=0$, the
system settles down with domains characterized by spins having this
reversed orientation destroying the free bosonic property. This
suggests that the random variable $\phi_{k_0}(0)$ attains the value
1 corresponding to $\theta_{k_0}=\pi$ as follows from
eqn.\eqref{eq4}. From this we find that the random variable
$\phi_{k_0}(\tau_q)\phi_{k_0}(0)$ acquires  the value 1 with
probability $p_{k_0}$. Thus we have
\begin{equation}
p_{k_0}=\langle \phi_{k_0}(\tau_q)\phi_{k_0}(0)\rangle, \label{eq22}
\end{equation}
and from eqn.\eqref{eq7} we find
\begin{equation}
p_{k_0}=\frac{1}{2}e^{-\omega_{k_0}\tau_q},  \label{eq23}
\end{equation}
$\omega_{k_0}$ being the frequency associated with the energy in the critical region.
\\
\noindent Now we note that the energy of the quasiparticle associated with a spin state
having the momentum mode $k_0$ is given by~\cite{18}
\begin{equation}
\epsilon_{k_0}=\lambda-2~cos~\frac{2\pi k_0}{N}, \label{eq24}
\end{equation}
$N$ being the number of sites. So taking the lattice constant
$a=\frac{2\pi}{N}=1$, we find at $\lambda=2$,
$\epsilon_{k_0}=2(1-cos~k_0)=4~sin^2\frac{k_0}{2} \sim k_0^2$ for
small $k_0$. This suggests that for $p_{k_0}$ we can write
\begin{equation}
p_{k_0}=\frac{1}{2}e^{-2\pi \tau_q k_0^2}. \label{eq25}
\end{equation}
The number density of defects (kinks) formed when the system finally settles down at $\lambda=0$ is given by
\begin{equation}
n_2=\frac{1}{2\pi}\int_{-\pi}^{\pi}p_{k_0}dk_0=\frac{1}{2\pi}.\frac{1}{2}\int_{-\pi}^{\pi}e^{-2\pi \tau_q k_0^2}dk_0=\frac{1}{4\pi}\frac{1}{\sqrt{2\tau_q}}. \label{eq26}
\end{equation}
Incidentally it has the same scaling behavior with $\tau_q$ as in
the transverse Ising model and the KZ correlation legth is
$\widehat{\xi} \sim \sqrt{\tau_q}$.
\\
\indent To estimate the entanglement entropy at criticality we note
that at $\lambda=0$ the system belongs to the free boson
universality class and the central charge of the relevant conformal
field theory corresponds to $c=1$. This gives the scaling law of the
entanglement entropy at $\lambda=0$ of a block of $L$ spins with the
rest of the system as
\begin{equation}
S_L \sim \frac{1}{3}~log_2L. \label{eq27}
\end{equation}
As mentioned earlier the entanglement entropy of a block of $L$
spins with the rest in a pure state is equivalent to the
entanglement of formation of a pair of nearest neighbor spins in a
mixed state given by concurrence $C$. As the concurrence $C$ is
associated with the Berry phase factor acquired by a spin state
while evolving in a closed path we write $C=|\phi|$ where the phase
is $e^{i2\pi \phi}$. Thus as in sec.\ref{sec2} using a RG flow
equation we can write for the entanglement entropy of a block of $L$
spins with the rest of the system
\begin{equation}
S_L \sim |\phi|~log_2L.  \label{eq28}
\end{equation}
At the critical point $\lambda=0$, the system belongs to the free
boson universality class, and the Berry phase factor $|\phi|$ here
is identical to the concurrence of a pair of nearest neighbor spins
in an antiferromagnetic system which is given by
$C=|\phi|=0.386$~\cite{6}. Now we take into account the correction
factor 0.926 due to the block spin renormalization. In fact the Von
Neumann entropy of a block of $L$ spins with the rest of the system
in pure state is here transcribed into the concurrence associated
with the entanglement of a renormalized block spin with another spin
corresponding to the rest of the system in a mixed state. This gives
the effective value of $|\phi|=0.386 \times 0.926=0.35$ which is
close to the value 1/3 derived from conformal field theory.
Utilizing this result we now estimate the entanglement entropy of a
block of $L$ spins with the rest of the system at criticality near
$\lambda=0$ induced by a quench. Indeed in analogy to
eqn.\eqref{eq17} we write
\begin{equation}
S(L,\tau_q)=4\frac{|\phi|}{|\phi|~ln~\tau_q}ln~L~\times 0.926=3.7\frac{ln~L}{ln~\tau_q}. \label{eq29}
\end{equation}
\begin{figure}[htbp]
\centering
\includegraphics[height=8cm,width=10cm]{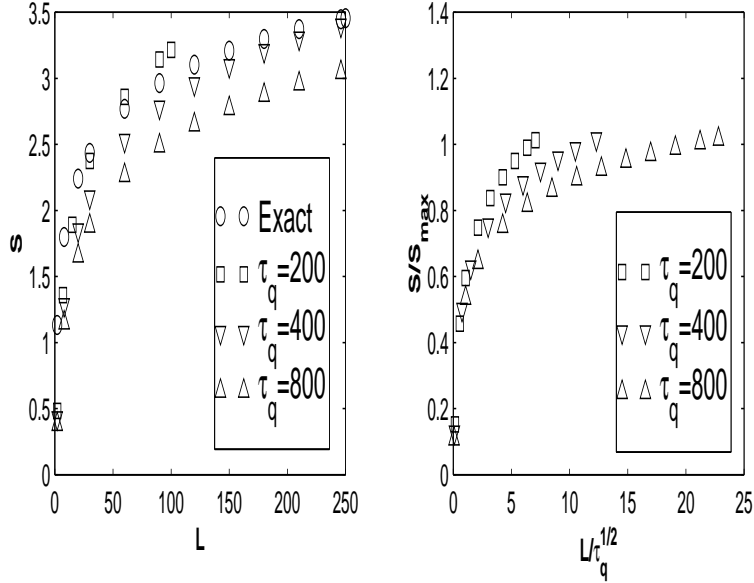}
\caption{\label{cap:figure_b} In fig.1 (left) the entanglement entropy of the XX model for various quench time ($\tau_q$) is compared with the exact value plotted for $\lambda=0$ from the result as reported in \cite{18}. Fig.1 (right) shows the variation of $S(L,\tau_q)/S_{max}$ with $L/\sqrt{\tau_q}$.}
\label{fig:fig_2}
\end{figure}
This result is independent of the prefactor in eqn.\eqref{eq28} and thus is universal for all one dimensional spin systems. However, the maximum value of $L$, $L_{max}$ allowed in the system is different. Following eqn.\eqref{eq18}, the maximum value of the entanglement entropy at $\lambda=0$ is derived as
\begin{eqnarray}
S_{max}&=& 2(|\phi|~log_2\widehat{\xi}+1)\times 0.926, \nonumber \\
&=& 0.25~ln~\tau_q+1.85.  \label{eq30}
\end{eqnarray}
Now from the constraint $S(L,\tau_q)/S_{max} \leq 1$ we obtain
\begin{equation}
ln~L \leq ~0.07~(ln~\tau_q)^2+0.5~ln~\tau_q.  \label{eq31}
\end{equation}
\\
\noindent Thus we find that due to generation of defects, the
scaling law for the entanglement entropy is restricted by the
maximum value of the block size allowed in the system. This picture is in the vicinity of the point $\lambda=0$. However, in the region $2>\lambda>0$, the entropy of
entanglement gradually decreases with the increase in the magnetic
field following the same scaling law until at $\lambda=2$, the
entanglement vanishes.
\\
It is pointed out by Latorre, Rico and Vidal~\cite{18a} that the
maximum entropy is reached at $\lambda=0$. The entropy is fitted by
a logarithmic scaling law as given by eqn.\eqref{eq27} with a
constant term which has been determined analytically by Jin and
Korepin~\cite{18b}. These results have been obtained without
critical slowing down when nonadiabaticity comes into play. It is
noted from eqn.\eqref{eq29} that when nonadiabaticity near
criticality is introduced the entropy decreases with quench time. In fig.\ref{fig:fig_2} (left) we have compared our result with the exact value
of the entropy at $\lambda=0$ as reported in \cite{18} (where no
quench was introduced) for $\tau_q=200,~400$ and $800$. As expected we
have found that the entropy increases with decreasing quench time.
At $\lambda=0$ the entropy is maximum and we find that our result is
in reasonable good agreement with the exact value for small value of
$\tau_q$. Indeed, we find that for $\tau_q=200$ it is close to the
exact value and for higher values of $\tau_q$ the entropy decreases. In fig.\ref{fig:fig_2} (right) we have plotted $S(L,\tau_q)/S_{max}$ vs. $L/\sqrt{\tau_q}$.

\vskip .5cm

\section{The XXX model} \label{sec4}

The Hamiltonian for the XXX model is given by
\begin{equation}
H_{XXX}=\sum_i\left(\sigma_i^x\sigma_{i+1}^x+\sigma_i^y\sigma_{i+1}^y+\sigma_i^z\sigma_{i+1}^z\right)+\sum_i\lambda \sigma_i^z. \label{eq32}
\end{equation}
The critical behavior of this model is analogous to that of the XX model. It has two limit
behavior. At $\lambda=2$ the system represents the ferromagnetic
state and at $\lambda=0$ the system corresponds to the
antiferromagnetic state. The interval $2>\lambda>0$ is gapless and
hence critical. Let us denote the time evolution
of the magnetic field by the relation  
\begin{equation}
\lambda(t<0)=-2t/\tau_q, \label{eq33}
\end{equation}
$\tau_q$ being the quench time and analyze the dynamics of the
system when it transits through the point $\lambda=2$. In the critical region, there is  random fluctuation in the Berry phase and from the two point correlation of the Berry
phase factor we derive the excitation probability.
Actually, this is given by the same relation as in eqn.\eqref{eq25} and we write
\begin{equation}
p_{k_0}=\frac{1}{2}e^{-\omega_{k_0}\tau_q}.  \label{eq34}
\end{equation}
In this case the energy of the quasiparticle in the
vicinity of the critical point is different from that of the XX
model.
\\
\noindent A generalization of the XXX model is the anisotropic Heisenberg model which is known as the XXZ model given by the Hamiltonian
\begin{equation}
H_{XXZ}=\sum_i\left(\sigma_i^x\sigma_{i+1}^x+\sigma_i^y\sigma_{i+1}^y+\Delta
\sigma_i^z\sigma_{i+1}^z\right)+\lambda \sum_{i}\sigma_i^z.
\label{eq35}
\end{equation}
It may be mentioned that for $\lambda=0$ the anisotropic system
shows a gapless phase in the interval $-1 \leq \Delta \leq 1$. For
$\Delta=1$ the system corresponds to the isotropic Heisenberg model.
For $\lambda>2$ the isotropic Heisenberg model corresponds to a
ferromagnetic state. As the magnetic field decreases and passes
through the critical point $\lambda=2$ some spins get flipped. For
an excited state of the Heisenberg model belonging to the subspace
of states where some spins are flipped the energy of the
quasiparticle with momentum mode $k_0$ is given by~\cite{19}
\begin{equation}
\epsilon_{k_0}=4(1-cos~k_0)=8sin^2~\frac{k_0}{2} \sim 2k_0^2, \label{eq39}
\end{equation}
for small $k_0$. Thus from the relation \eqref{eq34} we can write
\begin{equation}
p_{k_0}=\frac{1}{2}e^{-2\pi \epsilon_{k_0}\tau_q}=\frac{1}{2}e^{-4\pi \tau_qk_0^2}, \label{eq40}
\end{equation}
and the number density of defects is given by
\begin{equation}
n_3=\frac{1}{2\pi}\int_{-\pi}^{\pi}p_{k_0}dk_0=\frac{1}{4\pi}\int_{-\pi}^{\pi}e^{-4\pi \tau_q k_0^2}dk_0=\frac{1}{8\pi}\frac{1}{\sqrt{\tau_q}}.  \label{eq41}
\end{equation}
Thus the number density of kinks scales like $\tau_q^{-\frac{1}{2}}$
and the KZ correlation length  is given by $\widehat{\xi}\sim
\sqrt{\tau_q}$.
\\
\noindent As the XXX chain at $\lambda=0$ corresponds to the
antiferromagnetic state, the behavior of the entanglement entropy
for a block of $L$ spins with the rest of the system is identical
with that of the XX model. This is valid for $S(L,\tau_q)$ and
$S_{max}$ also. Just as in the XX model, in this model also, the
entanglement entropy is maximum for $\lambda=0$ which gradually
diminishes as the magnetic field increases, and finally it vanishes
at $\lambda=2$.

\vskip .5cm

\section{The Lipkin-Meshkov-Glick model} \label{sec5}

\noindent The Lipkin-Meshkov-Glick (LMG) model demonstrates the
mechanism of a phase transition for a many body system and was
introduced almost forty five years ago~\cite{20}. In contrast to the
conventional spin models in the LMG model each spin interacts with
all the spins of the system with the same coupling strength. This
highly symmetric interaction pattern introduces the loss of the
notion of geometry as there is no distance between the spins. This
implies that we cannot consider the notion of a block of spins as a
set of contiguous spins here. The symmetry of the Hamiltonian suggests that the ground state belongs to a symmetric subspace where all the spins are indistinguishable and this subspace restricts the entanglement entropy of a block of $L$ spins with the remaining spins. Interestingly the scaling behavior
shows a similar pattern with that of the XX model where conformal
symmetry plays a significant role at the critical region. But in the
LMG model the scaling law seems to have nothing to do with any
underlying conformal symmetry.
\\
\noindent In this note, we show that in case we introduce
point-splitting regularization the regularized LMG Hamiltonian in
the isotropic  case ($\gamma=1$) appears as a combination of
Heisenberg (XXX) model and Ising model with magnetic field along the
negative z-axis. In view of this, the regularized Hamiltonian
follows the pattern of the usual spin systems where we can conceive
of the notions of local interactions and translational invariance.
The underlying conformal symmetry at the critical region of these
systems is manifested in the LMG model and the similarity of the
pattern of the scaling behavior of the entanglement entropy of this
system can be understood. Thus the point-splitting regularization of
the LMG model uncovers the hidden conformal symmetry at criticality
which is lost when we take the sharp point limit.
\\
\noindent The Hamiltonian for the LMG model is given by
\begin{equation}
H=\frac{1}{N}\sum_{i<j}\left(\sigma_i^x\sigma_j^x+\gamma \sigma_i^y\sigma_j^y\right)+\lambda \sum_i\sigma_i^z. \label{eq42}
\end{equation}
$N$ being the total number of spins. 
\\
\noindent In what follows we shall consider the isotropic case with $\gamma=1$. The Hamiltonian \eqref{eq42} can be written in terms of the total spin operator $S^{\alpha}=\frac{1}{2}\sum \sigma_i^{\alpha}$ as
\begin{equation}
H=\frac{2}{N}\left(\vec{S}^2-(S^z)^2-\frac{N}{2}\right)+2\lambda S^z. \label{eq43}
\end{equation}
Let us consider that this total spin operator is located at a spatial point $k$. Now we introduce the point-splitting regularization so that we write
\begin{equation}
\vec{S}^2=\vec{S}_k\vec{S}_{k'}\delta_{kk'}, \label{eq44}
\end{equation}
where $k$ and $k'$ are two adjacent sites with an infinitesimal distance $k'-k=\epsilon$. Thus $\vec{S}^2$ corresponds to the product $\vec{S}_k\vec{S}_{k'}$ in the limit $\epsilon \rightarrow 0$. Taking
\begin{equation}
S_k^{\alpha}=\frac{1}{2}\sum_i\sigma_i^{\alpha},~~~~S_{k'}^{\alpha}=\frac{1}{2}\sum_j\sigma_j^{\alpha},\label{eq45}
\end{equation}
where $i$ and $j$ are two adjacent sites with an infinitesimal
distance.  Considering  only nearest neighbor interactions we can
write the Hamiltonian \eqref{eq43} in the regularized form
\begin{eqnarray}
H_{reg}&=&\frac{2}{N}\left[\frac{1}{4}\sum_{i,j}\left(\sigma_i^x\sigma_j^x+\sigma_i^y\sigma_j^y+\sigma_i^z\sigma_j^z\right)-
\frac{1}{4}\sum_{i,j}\sigma_i^z\sigma_j^z-\frac{N}{2}\right]+\lambda\sum_i\sigma_i^z, \nonumber \\
&=& \frac{1}{2N}\sum_{i,j}\left(\sigma_i^x\sigma_j^x+\sigma_i^y\sigma_j^y+
\sigma_i^z\sigma_j^z\right)-\frac{1}{2N}\sum_{i,j}\sigma_i^z\sigma_j^z-1+\lambda\sum_i\sigma_i^z,  \nonumber \\
&=&\frac{1}{2N}\left[\sum_{i,j}\left(\sigma_i^x\sigma_j^x+\sigma_i^y\sigma_j^y+\sigma_i^z\sigma_j^z\right)+N\lambda\sum_i\sigma_i^z\right]
-\frac{1}{2N}\left(\sum_{i,j}\sigma_i^z\sigma_j^z-N\lambda\sum_i\sigma_i^z\right) \nonumber \\
& &-1.  \label{eq46}
\end{eqnarray}
In deriving eqn.\eqref{eq46} we have considered nearest neighbor
interaction only because of the fact that the introduction of next nearest neighbor interaction gives rise to frustration in the
antiferromagnetic phase of the XXX model which is unsolicited in the LMG
model. In fact, this happens when the spin at the initial site
interacts with all odd sites. To avoid this unwanted feature only
the nearest neighbor interaction has been taken into account.
\\
\noindent Now considering $\frac{1}{2N}$ as a dimensionless coupling constant $J$ we write $H_{reg}=H_1+H_2-1$ with
\begin{eqnarray}
H_1 &=& J\left[\sum_{i,j}\left(\sigma_i^x\sigma_j^x+\sigma_i^y\sigma_j^y+\sigma_i^z\sigma_j^z\right)+
\frac{\lambda}{2J}\sum_i\sigma_i^z\right], \\
H_2 &=& -J\left(\sum_{i,j}\sigma_i^z\sigma_j^z-\frac{\lambda}{2J}\sum_i\sigma_i^z\right). \label{eq47}
\end{eqnarray}
The Hamiltonian given by eqn.\eqref{eq46} can be split into various other $H_1$ and $H_2$ such that both of them describe critical systems. The motivation behind this particular choice follows from the fact that cluster of even (odd) number of interacting spins represent bosonic (fermionic) systems. As the LMG model is characterized by the fact that each spin interacts with every other spin in the regularized Hamiltonian, we have chosen the critical systems such that one of these represents bosonic and the other fermionic features. In view of this we have taken $H_1$ and $H_2$ such that one represents the XXX model and the other Ising model.
\\
\noindent The criticality of this system can be studied by noting that this
is a combination of the XXX model given by $H_1$ and the Ising model
with magnetic field along the negative $z$ direction given by $H_2$.
It is observed that $\frac{\lambda}{2J}$ denotes the intensity of
the magnetic field. We know that the XXX model has two limit
behavior. When $\frac{|\lambda|}{2J}>2$, the XXX system is gapped
and for $\lambda=0$, the magnetization is zero and the system is in
an entangled (antiferromagnetic) state. In the interval
$2>\frac{|\lambda|}{2J}>0$ the system is gapless and this denotes
the critical region. Noting that the dimensionless parameter $J$
corresponds to $\frac{1}{2N},~N$ being an integer we find that
$\frac{|\lambda|}{2J}=2$ corresponds to the value
$|\lambda|=\frac{2}{N}$. Since in an entangled state the minimum
number of spins must be 2., i.e. $N_{min}=2$, at
criticality $|\lambda|$ lies in the interval $0<|\lambda|<1$. The value of $\lambda$ is independent of
the $real$ number of sites as we have treated the parameter
$\frac{1}{2N}$ as the coupling constant. Now it is observed that in the Ising chain given by $H_2$ for $|\lambda|>1$ the system is in the ferromagnetic state when all spins
are oriented along the negative z-axis. So in the interval
$0<|\lambda|<1$ as $|\lambda|$ is tuned from 1 to 0 the spin system
will undergo a transition when down spins will be excited and at
$\lambda=0$ all spins will settle down with opposite orientations.
Evidently, with the evolution of $|\lambda|$ the spins cross a point
when these are oriented along the x-axis. Hence, this corresponds to
the double ferromagnet transition and in the interval
$0<|\lambda|<1$ spins evolve through a paramagnetic state. Thus
during this transition in the interval $0<|\lambda|<1$ the
entanglement entropy of the spins evolves through a nonzero value
which is similar to that of the transverse Ising model. In the
regularized Hamiltonian we can consider the time dependent magnetic
field corresponding to the summation of that given by eqns.\eqref{eq3}
and \eqref{eq33} and so we write
\begin{equation}
\lambda(t<0)=-\frac{2t}{\tau_q}+\frac{t}{\tau_q}=-\frac{t}{\tau_q}, \label{eq48}
\end{equation}
$\tau_q$ being the quench time. We have taken $\frac{t}{\tau_q}$ for
the Ising system with positive sign as the magnetic field here is in
the negative z-direction. Finally, when the system settles down at $t=0~~(\lambda=0)$,  the number density of defects is estimated and is given by $n_4=n_1+n_3$ where $n_1~(n_3)$ is given by
eqns.\eqref{eq11} (\eqref{eq41}) in the thermodynamic limit. From this we obtain
\begin{equation}
n_4=n_1+n_3=\frac{1}{2\pi}\frac{1}{\sqrt{2\tau_q}}+\frac{1}{8\pi}\frac{1}{\sqrt{\tau_q}}=\frac{2\sqrt{2}+1}{8\pi\sqrt{\tau_q}}.  \label{eq49}
\end{equation}
Thus we note that in LMG model also the number density of kinks formed
scales as $\tau_q^{-\frac{1}{2}}$. However, in this model a
correlation length characterizing the typical distance between
defects cannot be introduced  though one can estimate the
fraction of flipped spins after the quench.
\\
\noindent The entanglement entropy in the LMG model of a block of $L$ spins with respect to the rest of the spins can be derived in an analogous way. The entanglement entropy in the critical region $0<|\lambda|<1$ is given by the summation of the entanglement entropy of the XXX model and the transverse Ising model at criticality. As the entanglement entropy for a block of $L$ spins with the rest of the system is given by $S_1(L)\sim \frac{1}{3}~log_2L$ for the XXX model and that for the transverse Ising model is given by $S_2(L)\sim \frac{1}{6}~log_2L$,  the entanglement entropy in the critical region $0<|\lambda|<1$ for the LMG model corresponds to the scaling law in the thermodynamical limit as
\begin{equation}
S(L)\sim \frac{1}{3}~log_2L+\frac{1}{6}~log_2L \sim \frac{1}{2}~log_2L,  \label{eq50}
\end{equation}
which is identical with that obtained by Latorre, Orus, Rico and Vidal~\cite{21}. One should note that for $\lambda \geq 1$, the system represents a polarized product state. Thus when the LMG model is recast in the regularized form the critical region manifests the same logarithmic scaling law as observed in XXX model and the transverse Ising model which are characterized by conformal symmetry at the critical region in these systems. Though the LMG model appears to have nothing to do with conformal symmetry at the critical region, the point-splitting regularization reveals an underlying conformal symmetry at criticality in this system which is lost in the sharp point limit.
\\
\indent Now we consider that the criticality is induced by a linear quench \eqref{eq48}. As the regularized Hamiltonian suggests that at the critical region this corresponds to a combination of the XXX model and the transverse Ising model so following eqns.\eqref{eq17} and \eqref{eq29}, we write
\begin{equation}
S(L,\tau_q)=\frac{2|\widetilde{\phi}|~log_2L}{|\widetilde{\phi}|~log_2\widehat{\xi}}\times 0.926, \label{eq51}
\end{equation}
where $|\widetilde{\phi}|=|\phi|_{XXX}+|\phi|_{Ising}=0.386+0.18=0.566$ and $\widehat{\xi}~(\sim \sqrt{\tau_q})$ is the associated KZ correlation length. The final result is independent of $|\widetilde{\phi}|$ and we obtain the universal value
\begin{equation}
S(L,\tau_q)=3.7\frac{ln~L}{ln~\tau_q}. \label{eq52}
\end{equation}
Here the maximum value of $L$, $L_{max}$ is different from that of the XXX and the transverse Ising model. In analogy to eqns.\eqref{eq18} and \eqref{eq30} we find that the maximum value of the entanglement entropy
\begin{equation}
S_{max}|_{LMG}=2(|\widetilde{\phi}|~log_2\widehat{\xi}+1)\times 0.926 \approx 0.36~ln~\tau_q+1.85.    \label{eq53}
\end{equation}
So from the constraint $S(L,\tau_q)/S_{max}\leq 1$ we find
\begin{equation}
ln~L|_{LMG} \leq 0.097~(ln~\tau_q)^2+0.5~ln~\tau_q.  \label{eq54}
\end{equation}
\\
\noindent In the sharp point limit the LMG model does not allow any correlation length. However, we can consider the coherence number~\cite{22} which can be identified with the KZ correlation length $\widehat{\xi}$ in the regularized formalism.
\begin{figure}[htbp]
\centering
\includegraphics[height=6cm,width=12cm]{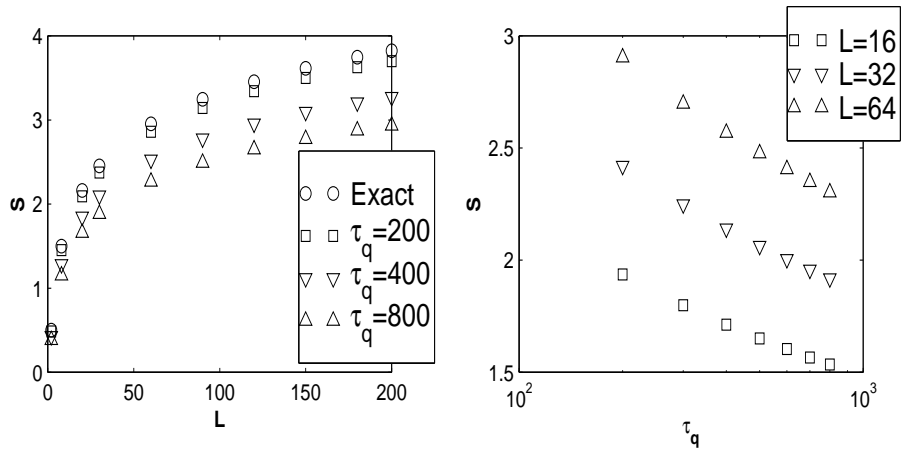}
\caption{\label{cap:figure_c} In fig.2 (left) we have compared our result for $S(L,\tau_q)$ for $\tau_q=200,~400$ and 800 with the values derived from that reported in \cite{18} at $\lambda=0$ extrapolating it to the thermodynamic limit. In fig.2 (right) we have plotted our results for $L=16,~32$ and 64 in the region $\tau_q > 200$.}
\label{fig:fig_3}
\end{figure}
It is noted that in the critical region $0<\lambda<1$ the entanglement entropy is maximum at $\lambda=0$ and decreases with the increase in the magnetic field until at $\lambda=1$ it vanishes when the system represents a product state. However, within this region the scaling behavior remains the same.
\\
\noindent It may be pointed out here that recently the LMG model has
been studied in which the system is dragged adiabatically through
the critical point~\cite{23}. Using the Landau-Zener
formula~\cite{24} the fraction of the flipped spins has been found
to scale like $\tau_q^{-\frac{3}{2}}$, $\tau_q$ being the quench
time. The failure of obtaining the scaling behavior
$\tau_q^{-\frac{1}{2}}$ as derived here may be related to the
definition of the defect density in this formalism. Indeed, here the
degree of adiabaticity is estimated through the residual energy
$E_{res}$ given by $E_{res}=E_{fin}-E_{gs}$ where $E_{fin}$ is the
average energy in the final time-evolved state and $E_{gs}$ is the
ground state energy. 
\begin{figure}[htbp]
\centering
\includegraphics[height=6cm,width=6cm]{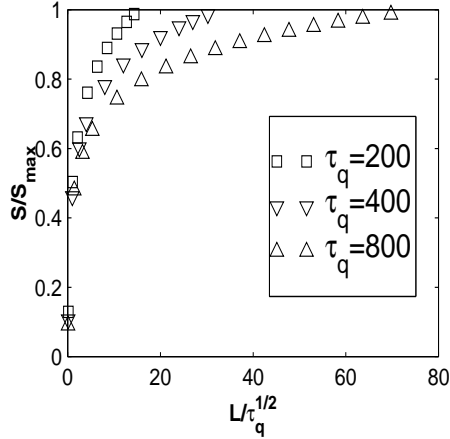}
\caption{\label{cap:figure_c_1}The entanglement entropy $S(L,\tau_q)$ vs. $L/\sqrt{\tau_q}$.} 
\label{fig:fig_3_a}
\end{figure}
It is found that the system has three regions:
(i) for fast quenches $E_{res}$ persists independent of $\tau_q$,
(ii) for slower quenches $E_{res}$ persists $\sim
\tau_q^{-\frac{3}{2}}$ and (iii) for further slowing $E_{res}$
persists $\sim \tau_q^{-2}$. It is noted that for fast quench, the
transition is nonadiabatic. However, for slow transitions the
adiabatic dynamics is quite different from our present formalism. In
fact, the result we have obtained here is fully determined by the
nonadiabaticity condition in the vicinity of the critical point.
\\
\noindent In fig.\ref{fig:fig_3} (left) we have compared our result for the entanglement entropy for the LMG model with the value for $\lambda=0$ derived in \cite{18} extrapolating it to the thermodynamic limit. It is noted that when we introduce a quench the entropy decreases with the increase in the quench time $\tau_q$. So it is expected that the value reported in \cite{18} will be in agreement with our result for small $\tau_q$. However for very small $\tau_q$, the maximum value of the block size $L$ will be very small which follows from eqn.\eqref{eq54}. In view of this we have taken moderate values of $\tau_q=200,~400$ and 800 for comparison. It is noted that for $\tau_q=200$ the result is very close to the value derived from \cite{18}.
\\
\noindent Caneva \textit{et. al.}~\cite{23} have computed the entanglement entropy at $\lambda=0$ for various values of $\tau_q$ when the system is adiabatically drugged through the critical point for finite system size. According to their result, for fast quench $\tau_q \rightarrow 0$ the entanglement entropy tends to be vanishing and for slow dynamics  $\tau_q \rightarrow \infty$  the entropy picks up the value it assumes in the final ground state $S_{gs}~(\lambda=0)=1$ independent of the system size. Between this two limiting behaviors the entropy reaches a size dependent maximum at $\tau_q$ nearing 10 and then decreases with the increase in $\tau_q$. In our formalism we find that for $\tau_q$ nearing 10, the maximum value of the block size $L$ will be very small as follows from eqn.\eqref{eq54}. In view of this we have taken moderately large $\tau_q > 200$ for comparison with their result. Caneva \textit{et. al.}~\cite{23} have computed their result for block size $L=N/2$, $N$ being the system size. However our result has been computed in the thermodynamic limit. In fig.\ref{fig:fig_3} (right) we have plotted our results for $L=16,~32$ and 64 and $\tau_q > 200$. We find that our result is in reasonable good agreement with that of Caneva \textit{et. al.}~\cite{23}. In fig.\ref{fig:fig_3_a} we have plotted $S(L,\tau_q)$ vs. $L/\sqrt{\tau_q}$.      

\vskip .5cm

\section{Defect formation and maximal entanglement entropy in different one dimensional systems: A comparative study} \label{sec6}

\noindent We have studied here defect formation and the scaling of
entanglement entropy in QPT in different one
\begin{figure}[htbp]
\centering
\includegraphics[height=8cm,width=12cm]{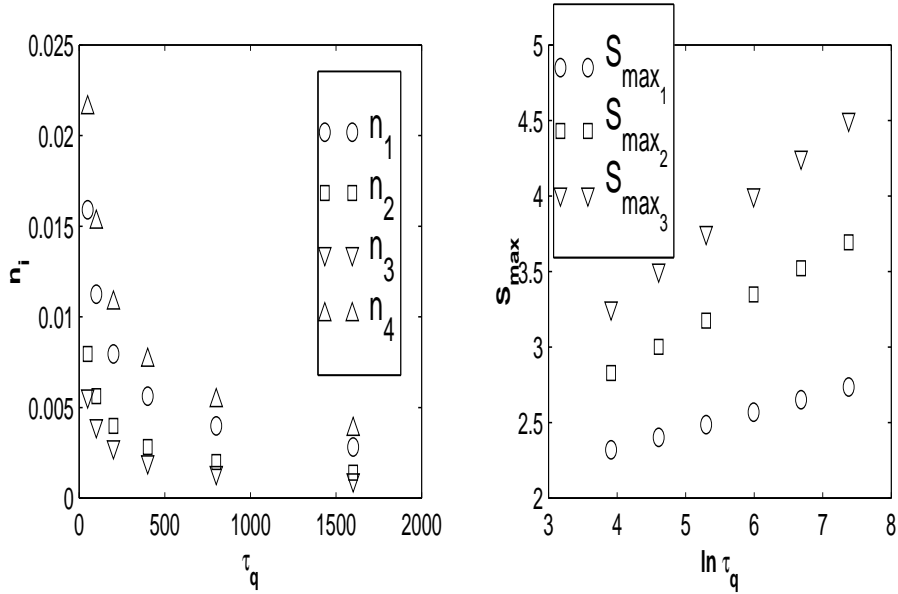}
\caption{\label{cap:figure_d}Fig.4 (left) shows the variation of the number density of defects $n_1,~n_2,~n_3,~n_4$ with quench time $\tau_q$ in isotropic XY, XX, XXX and LMG models respectively for different values of $\tau_q$. Fig.4 (right) shows the maximum value of the entanglement entropy $S_{max}$ for different values of $\tau_q$ in isotropic XY ($S_{{max}_1}$), XX (XXX) ($S_{{max}_2}$) and LMG ($S_{{max}_3}$) models. It is noted that $S_{max}$ is identical for XX and XXX systems.}
\label{fig:fig_4}
\end{figure}
dimensional systems induced by a quench. It is found that they all
show a universal behavior. The number density of defects in all
these systems scales like $\tau_q^{-\frac{1}{2}}$ with difference
only in the prefactors. Also the entanglement entropy at criticality
takes a universal value having the same scaling pattern with block
spin size $L$. However, there is a restriction on the maximum value
of $L$ allowed in different systems. The maximum value of the
entropy in all these systems follows the same pattern with
difference only in the coefficients. In all these systems the
constraint on the maximum value of the block size $L$ restricts the
validity of the scaling law for any arbitrary $L$. In fact, the main
implication of the scaling law for the entanglement entropy is that
with the addition of every spin, entanglement increases. However, in
QPT induced by a quench this is not valid beyond the upper limit of
the value of $L$ allowed in various systems. Indeed, the
nonadiabatic transition due to quench makes the scaling law a
restrictive one and cannot be treated in the conventional sense.
\\
\noindent In fig.\ref{fig:fig_4} (left) we have plotted the number density of defects formed during QPT induced by a quench in different spin systems for different values of $\tau_q$. The density of defects for the isotropic XY, XX, XXX, LMG models are denoted by $n_1,~n_2,~n_3,~n_4$ respectively. In fig.\ref{fig:fig_4} (right) we have plotted the maximum value of entanglement entropy $S_{max}$ for different values of $\tau_q$ in isotropic XY, XX and LMG models. $S_{max}$ is identical for XX and XXX systems.

\vskip .5cm

\section{Discussion} \label{sec7}

We have analyzed here the defect formation and the scaling law of
the entanglement entropy in QPT in several one-dimensional systems
at critical slowing down when nonadiabaticity plays  a dominant role.
We have argued that in the vicinity of the critical point
nonadiabaticity in QPT causes spin fluctuation which in turn makes
the Berry phase factor a random one. From the two-point correlation
of the Berry phase factor we have estimated the probability of the
generation of defects. Indeed, in a recent paper~\cite{10} we have
estimated the number density of defects as well as the spin-spin
correlation at criticality of the transverse Ising model from this
dynamical mechanism of QPT induced by a quench. The results are
found to be identical with those derived from the standard
Landau-Zener transition probability studied by other
authors~\cite{3,4,5}. Here we have analyzed several other one dimensional spin
systems using this dynamical mechanism. Our analysis suggests that
this formalism  represents the universal dynamics of
QPT induced by a quench.


\begin{thebibliography}{}
\bibitem{17} There is a typographical error in eqn.(26) in \cite{11}. The correct prefactor will be 0.03 and not 0.3 as printed.
\end{thebibliography}
\end{document}